\documentclass[%
 aip,
 amsmath,amssymb,
 reprint,%
]{revtex4-1}

\usepackage{graphicx,color}
\usepackage{dcolumn}
\usepackage{bm}

\usepackage[utf8]{inputenc}
\usepackage[T1]{fontenc}
\usepackage{mathptmx}
\usepackage{etoolbox}

\usepackage{color}
\usepackage{textcase, url}
\usepackage[normalem]{ulem}
\usepackage{siunitx}

\definecolor{darkblue}{rgb}{0,0,0.5}
\definecolor{darkgreen}{rgb}{0,0.5,0}
\definecolor{darkred}{rgb}{.7,0,0}
\definecolor{purple}{rgb}{0.5,0,0.6}
\definecolor{orange}{rgb}{1,0.5,0}
\definecolor{grey}{rgb}{.6,.6,.6}
\definecolor{lightpink}{rgb}{1,0.7,0.75}
\definecolor{pink}{rgb}{1,0.4,0.58}
\definecolor{deeppink}{rgb}{1,0.08,0.58}
\definecolor{brown}{rgb}{0.59, 0.29, 0.0}
\definecolor{blue-green}{rgb}{0.0, 0.87, 0.87}

\makeatletter
\def\@email#1#2{%
 \endgroup
 \patchcmd{\titleblock@produce}
  {\frontmatter@RRAPformat}
  {\frontmatter@RRAPformat{\produce@RRAP{*#1\href{mailto:#2}{#2}}}\frontmatter@RRAPformat}
  {}{}
}%
\makeatother
\begin{document}

\preprint{AIP/123-QED}

\title{Generation of a single-cycle surface acoustic wave pulse on LiNbO$_3$ for application to thin film materials}
\author{Koji Fujiwara}
\affiliation{Department of Physics, Graduate School of Science, Osaka University, Toyonaka, Osaka 560-0043, Japan}

\author{Shunsuke Ota}
\affiliation{Department of Electrical and Electronic Engineering, Institute of Science Tokyo, Meguro-ku, Tokyo 152-8552, Japan}
\affiliation{National Metrology Institute of Japan (NMIJ), National Institute of Advanced Industrial Science and Technology (AIST), Tsukuba, Ibaraki 305-8563, Japan}

\author{Tetsuo Kodera}
\affiliation{Department of Electrical and Electronic Engineering, Institute of Science Tokyo, Meguro-ku, Tokyo 152-8552, Japan}

\author{Yuma Okazaki}
\affiliation{National Metrology Institute of Japan (NMIJ), National Institute of Advanced Industrial Science and Technology (AIST), Tsukuba, Ibaraki 305-8563, Japan}

\author{Nobu-Hisa Kaneko}
\affiliation{National Metrology Institute of Japan (NMIJ), National Institute of Advanced Industrial Science and Technology (AIST), Tsukuba, Ibaraki 305-8563, Japan}

\author{Nan Jiang}
\affiliation{Department of Physics, Graduate School of Science, Osaka University, Toyonaka, Osaka 560-0043, Japan}
\affiliation{Center for Spintronics Research Network, Osaka University, Toyonaka, Osaka 560-8531, Japan}
\affiliation{Institute for Open and Transdisciplinary Research Initiatives, Osaka University, Osaka 565-0871, Japan}

\author{Yasuhiro Niimi}
\affiliation{Department of Physics, Graduate School of Science, Osaka University, Toyonaka, Osaka 560-0043, Japan}
\affiliation{Center for Spintronics Research Network, Osaka University, Toyonaka, Osaka 560-8531, Japan}
\affiliation{Institute for Open and Transdisciplinary Research Initiatives, Osaka University, Osaka 565-0871, Japan}

\author{Shintaro Takada}
\affiliation{Department of Physics, Graduate School of Science, Osaka University, Toyonaka, Osaka 560-0043, Japan}
\affiliation{Center for Spintronics Research Network, Osaka University, Toyonaka, Osaka 560-8531, Japan}
\affiliation{Institute for Open and Transdisciplinary Research Initiatives, Osaka University, Osaka 565-0871, Japan}
\affiliation{Center for Quantum Information and Quantum Biology (QIQB), Osaka University, Osaka 565-0871, Japan}
\email{takada@phys.sci.osaka-u.ac.jp}

\date{\today}

\begin{abstract}
Surface acoustic wave (SAW) technology has been explored in thin-film materials to discover fundamental phenomena and to investigate their physical properties.
It is used to excite and manipulate quasi-particles such as phonons or magnons, and can dynamically modulate the properties of the materials.
In the field, SAWs are typically excited by a continuous wave at a resonant frequency.
Recently, generation of a single-cycle SAW pulse has been demonstrated on GaAs substrate.
Such a SAW pulse provides a potential to access a single quasi-particle excitation and to investigate its dynamics by time-resolved measurements.
On the other hand, to modulate and control the properties of thin film materials, it is generally required to generate high-intensity SAWs.
In this work, we demonstrate the efficient generation of a SAW pulse using a chirp interdigital transducer (IDT) on LiNbO$_3$ substrate.
We have fabricated chirp IDT devices with bandwidths from \SI{0.5}{GHz} to \SI{5.5}{GHz}.
We also confirmed the generation of a SAW pulse with \SI{0.3}{ns} FWHM (full width at half maximum) by performing time-resolved measurements.
The conversion efficiency between input power and SAW on LiNbO$_3$ substrate is approximately 45 times larger than that on GaAs substrate.
This enables us to generate a high-intensity SAW pulse, meeting the requirement for the modulation of thin films.
Our results will expand the research in the field, such as spintronics and magnonics, and lead to their further advancements.
\end{abstract}
\maketitle

Surface acoustic waves (SAWs) are sound waves that propagate along solid surfaces.
SAW-based technologies are utilized in many electronic devices, such as microwave bandpass filters and various sensors for detecting pressure, temperature, and chemical vapors, among others.
SAWs have also found applications in various fields of fundamental research, inclusing superconducting quantum circuits, semiconductor electron systems, and spintronics \cite{Delsing_2019, Nie_2023, Zhao_2024}.

In general, SAWs are electrically generated and detected by using an interdigital transducer (IDT) fabricated on a piezoelectric substrate \cite{White_1965}.
Piezoelectric effect converts electrical signals into mechanical vibrations and vice versa.
Due to piezoelectric coupling, SAWs transmit both mechanical strain and electric fields.
The piezoelectric material is chosen depending on the purpose of the research.
GaAs substrate is, for example, employed to demonstrate quantized charge pumping \cite{Shilton_1996}, single-electron transport \cite{Hermelin_2011, McNeil_2011}, exciton transport \cite{Rudolph_2007}, or single-photon emission \cite{Hsiao_2020, Yuan_2021}.
In these experiments, GaAs substrate is employed since it can host high mobility two-dimensional electron gas.
On the other hand, GaAs has a relatively small electromechanical coupling factor $K^2$ that describes the efficiency of piezoelectric coupling, about \SI{0.07}{\%}.

For more efficient generation and detection of SAWs, LiNbO$_3$ is a typical choice of a substrate.
$K^2$ of 128$^{\circ}$ Y-cut LiNbO$_3$ is about \SI{5.6}{\%}, which is 80 times larger than GaAs.
Over the past decade, thin film materials have been deposited on a LiNbO$_3$ substrate and electric fields and/or mechanical strain accompanying to SAWs are utilized to modulate their electronic properties \cite{Nie_2023, Zhao_2024}.
In particular, van der Waals materials \cite{Novoselov_2004} have been intensively studied
and a variety of physics has been investigated \cite{Fandan_2020, Yokoi_2020, Zhao_2022, Fang_2023, Lyons_2023}.
SAWs have also been employed in spintronics, where, for example, switching magnetization \cite{Davis_2010, Foerster_2017} and generation of spin current \cite{Weiler_2012, Kobayashi_2017} or skyrmions \cite{Yokouchi_2020} have been reported.
In the applications, the amplitude of the modulations induced in the thin films can be lowered compared to the one at the surface of the piezoelectric substrate without any thin films on top.
For example, the electric field inside a metallic thin film induced by SAWs reduces several orders of magnitude.
As a result, it is critical to employ a piezoelectric substrate with a large $K^2$ such as LiNbO$_3$ to efficiently modulate the electronic properties of materials deposited on top.

\begin{figure*}
\includegraphics[width=17cm]{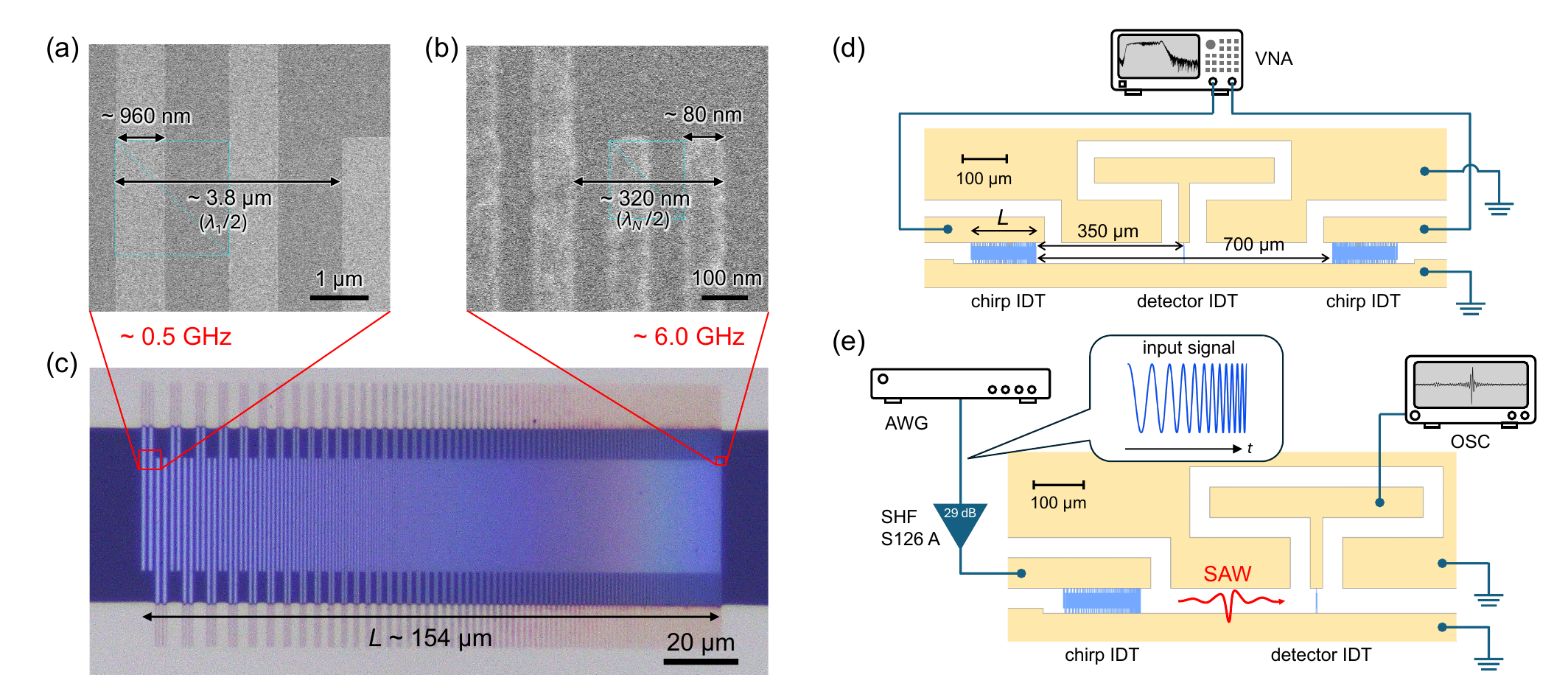}
\caption{\label{fig:dev_pic}
(a), (b) SEM image of the fabricated chirp IDT. The region zoomed in (a) is for $f_1 \approx \SI{0.5}{\GHz}$ SAW excitation and the one in (b) is for $f_N \approx \SI{6.0}{\GHz}$ SAW excitation.
(c) Optical microscope image of the whole chirp IDT ($T$ = \SI{40}{\ns}). The length of the chirp IDT $L$ is given by: $L = T v_\mathrm{SAW(design)}$.
(d) Schematic illustration of the whole device and setup used in frequency-response measurement between two chirp IDTs. Two chirp IDTs are placed separated by \SI{700}{\um}. Their orientations are reversed between left and right. The detector IDT is placed at the middle of the two chirp IDTs. Transmission property is also measured between the chirp IDT and the detector IDT.
(e) Schematic illustration of the setup used in time-resolved measurement.}
\end{figure*}

Typically, a SAW used in the research introduced above is a sinusoidal wave at the resonant frequency of the IDT having a constant periodicity.
On the other hand, it is also possible to use, for example, a rectangle or a saw-tooth wave of SAW by employing a proper IDT structure \cite{SchuleinFlorian_2015}.
SAWs propagate while keeping their shape due to their linear dispersion.
Recently, the generation of a single cycle-SAW pulse has been demonstrated by employing the IDT with a chirped structure, so-called a chirp IDT, on GaAs substrate \cite{Wang_2022}.
In the work, the SAW pulse was applied to demonstrate high efficiency single-electron transfer.
When the SAW pulse is applied to thin-film materials, it will expand the associated research field
and open the possibility, for example, to study the dynamics of SAW induced phenomena.
However, as discussed above such applications generally requires high intensity SAWs
and hence a substrate having a large $K^2$ is employed for efficient generation of SAWs.
Therefore, it is desirable to demonstrate the generation of a single-cycle SAW pulse
on a substrate having a larger $K^2$ than a GaAs substrate.

In this work, we fabricate chirp IDT devices on 128$^{\circ}$ Y-cut LiNbO$_3$ substrate for the application to modulate the electronic properties of thin film materials deposited on top.
We measure the frequency response of the devices with different numbers of IDT finger pairs. 
We, then, demonstrate the generation of a SAW pulse by time-resolved measurements.
The amplitude of the SAW signal is expected to increase dramatically compared to GaAs according to high $K^2$ factor of LiNbO$_3$.

SAWs employed to modulate thin-film materials are typically generated by an IDT having a certain number of finger pairs fabricated with a constant period, $\lambda$.
When the IDT is excited by a continuous radio-frequency signal with a frequency of $f = v_\mathrm{SAW}/\lambda$ where $v_\mathrm{SAW}$ is a velocity of SAWs, SAWs excited from each finger pair constructively interfere.
As a result, higher intensity SAWs are generated continuously.
Here we employed a chirp IDT \cite{Wang_2022} to generate a single-cycle SAW pulse.
The finger spacing of the chirp IDT gradually varies from $\lambda_1$ to $\lambda_N$, allowing each pair of fingers to excite SAWs with different frequencies.
By applying an appropriate voltage to the chirp IDT with a frequency varying from $f_1$ to $f_N$, SAWs are superposed in the bandwidth from $f_1$ to $f_N$.
As SAWs have linear dispersion over a wide frequency band, the waveform of the SAW pulse remains unchanged while it propagates.
In-phase superposition of SAWs with different frequencies can generate a single-cycle SAW pulse, similar to the generation of a delta function.

When designing a chirp IDT, parameters $v_\mathrm{SAW(design)}$, $T$, $f_1$ and $f_N$ are determined first. $v_\mathrm{SAW(design)}$ is the speed of SAWs in the area of the IDT, $T$ is the time connected to the total spatial length of the IDT as $L=Tv_\mathrm{SAW(design)}$. $f_1$ and $f_N$ are the minimum and the maximum frequencies of SAWs generated by the IDT.
Here $v_\mathrm{SAW(design)} = \SI{3800}{m/s}$ was employed taking into account the damping of SAWs by metal fingers of the IDT.
We chose $f_1$ and $f_N$ to be \SI{0.5}{GHz} and \SI{3}{GHz}, where the maximum and the minimum wavelengths of SAWs, $\lambda_n = v_\mathrm{SAW(design)/f_n}$, became \SI{7.5}{\um}, \SI{1.27}{\um}, respectively.
The number of finger pairs $N$ was determined by satisfying $\sum_{n=1}^N \lambda_n /v_\mathrm{SAW(design)} \approx T$.
For $T = \SI{40}{\ns}$, $N$ is 56 and $L$ becomes about \SI{154}{\um}.
We also fabricated chirp IDTs designed with $f_1 = \SI{0.5}{\GHz}$ and $f_N = \SI{6.0}{\GHz}$ ($\lambda _N \approx \SI{0.63}{\um}$).
In addition, we fabricated a broadband-detector IDT between two chirp IDTs to evaluate the SAW shape.
The full width at half maximum (FWHM) of the IDT resonance is given by $ f_\mathrm{0d} / N_\mathrm{pd}$, where $f_\mathrm{0d}$ is the resonant frequency and $N_\mathrm{pd}$ is a number of finger pairs of the detector IDT.
As a result, a smaller $N_\mathrm{pd}$ enables detection of a wide band SAW signals.
In this study, the detector IDTs were designed with $N_\mathrm{pd} = 1.5$ for all devices,
$f_\mathrm{0d} = \SI{2.5}{\GHz}$ ($\lambda_\mathrm{0d} = \SI{1.52}{\um}$) for \SI{3}{\GHz} devices and 
$f_\mathrm{0d} = \SI{4.5}{\GHz}$ ($\lambda_\mathrm{0d} = \SI{0.844}{\um}$) for \SI{6}{\GHz} devices.
The length of the aperture was chosen to be \SI{30}{\micro m} for all the IDTs.

We used open-source Python library ``idtpy''~\cite{idtpy_junliang} to design and simulate IDTs.
The electrode of IDTs were fabricated using electron-beam lithography with \SI{3}{\%} polymethyl-methacrylate (PMMA) resist.
Ti (\SI{3}{\nm}) and Al (\SI{27}{\nm}) were deposited onto the substrate in a vacuum chamber.
We also fabricated the contact pads using laser lithography and Ti (\SI{20}{\nm}) and Au (\SI{80}{\nm}) were deposited.
Scanning electron microscope (SEM) and optical microscope images of a typical chirp IDT device are shown in Figs.~\ref{fig:dev_pic}(a), \ref{fig:dev_pic}(b) and \ref{fig:dev_pic}(c).
The chirp IDT used in this study adopts a double-finger configuration, as also employed in Ref.~\onlinecite{Wang_2022}.
Such structures are typically used to suppress acoustic resonance within the IDT~\cite{Bristol_1972_double_finger, DeVries_1972_double_finger}.

Figure~\ref{fig:dev_pic}(d) and \ref{fig:dev_pic}(e) are schematic illustrations of the measurement setup
for characterization of frequency response and time-resolved measurement, respectively.
We measured the transmission $S$-parameter ($S_{21}$) to characterize the frequency response
of IDTs using the vector network analyzer (Keysight E5080A).
For time-resolved measurement, we used the arbitrary waveform generator (AWG, Keysight M8190A)
to apply a chirped signal to a chirp IDT.
In order to characterize the shape of SAWs, the sampling oscilloscope (OSC, Keysight N1010A)
was connected to a detector IDT.
The broadband amplifier (SHF S126 A, \SI{80}{kHz} - \SI{25}{GHz}, Gain \SI{29}{dB})
was inserted between AWG and a chirp IDT to improve the signal-to-noise ratio.
In this study, all measurements were performed by a semiautomatic-probe system at room temperature.

\begin{figure}
\includegraphics[width=8.5cm]{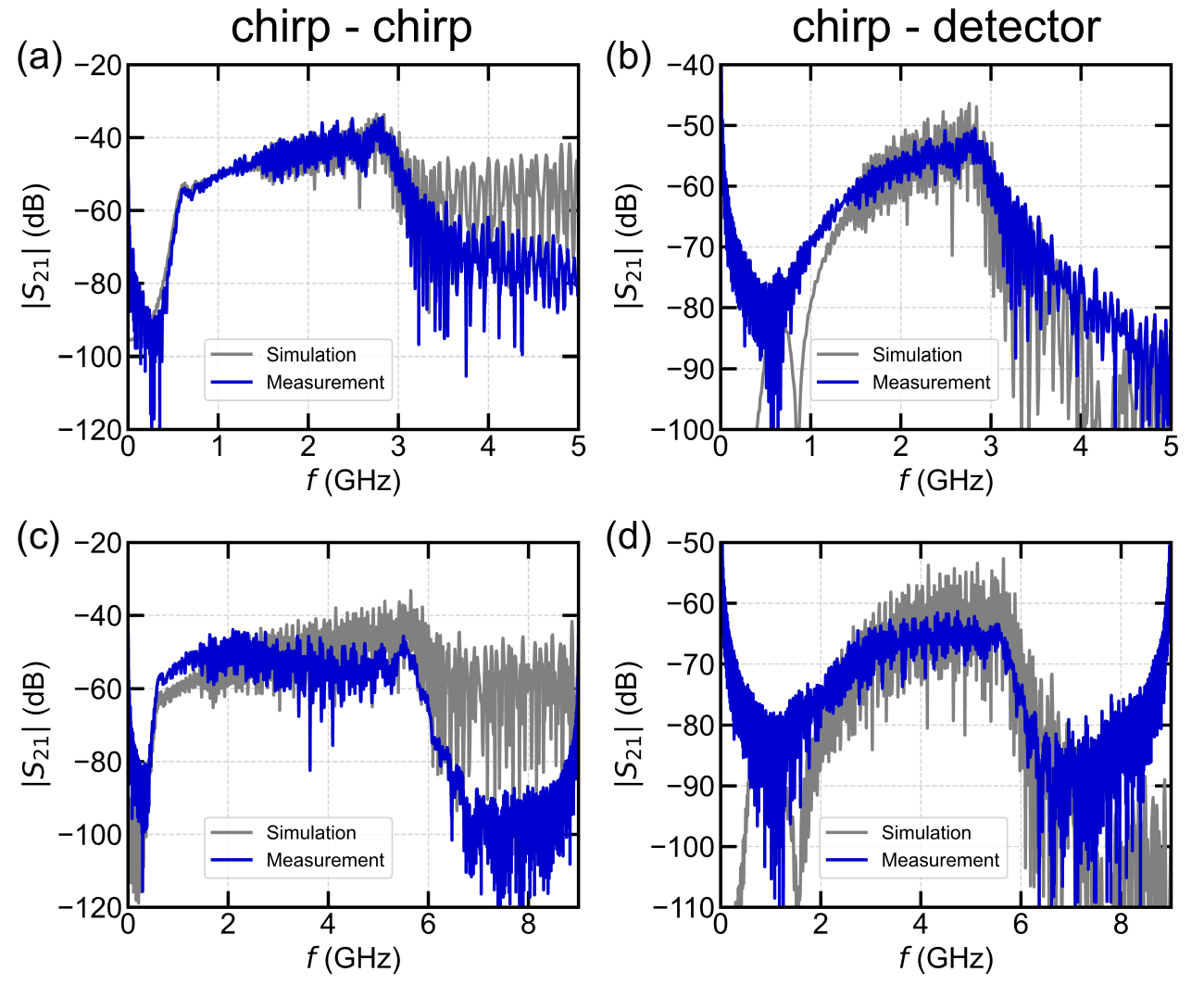}
\caption{\label{fig:Spara_timegate}
(a) Transmission property $|S_{21}|$ between a pair of chirp IDTs with
$T = \SI{40}{ns}$, $f_1 = \SI{0.5}{\GHz}$, $f_N = \SI{3.0}{\GHz}$, and $v_\mathrm{SAW(design)} = \SI{3800}{m/s}$.
These data were obtained by time-domain gating analysis to exclude electromagnetic crosstalk components.
The gray line shows the simulated response of the designed chirp IDT based on delta-function model. 
(b) Transmission property $|S_{21}|$ between the chirp IDT to the detector IDT with 1.5 pairs of fingers, $f_{\rm 0d} = \SI{2.5}{GHz}$, $v_{\rm SAW(design)} = \SI{3800}{m/s}$. The gray line is obtained by convoluting the simulated response of the chirp IDT and the detector IDT.
(c), (d) The same datasets as (a), (b) for the devices designed with $f_1 = \SI{0.5}{GHz}$, $f_N = \SI{6.0}{GHz}$. The detector IDT is a regular IDT with 1.5 pairs of fingers, $f_\mathrm{0d} = \SI{4.5}{GHz}$, $v_\mathrm{SAW(design)} = \SI{3800}{m/s}$.
}
\end{figure}

Firstly, we measured frequency response between two chirp IDTs to characterize the device properties.
The transmission property of the chirp IDTs designed with $T = \SI{40}{\ns}$,
$f_N = \SI{3.0}{GHz}$
measured by VNA is shown with the blue line in Fig.~\ref{fig:Spara_timegate}(a).
The result shows that power transmission by SAWs appears from around \SI{0.5}{GHz} to \SI{3.0}{GHz}.
The average value of $|S_{21}|$ between $f = \SI{0.5}{GHz}$ and \SI{3.0}{GHz} was \SI{-46}{dB}.
This is approximately \SI{30}{dB} larger than the one on GaAs~\cite{Wang_2022},
and hence 32 times higher power conversion efficiency
at each chirp IDT is expected.
Similar measurements were performed on a device designed with
$f_N = \SI{6.0}{GHz}$ and $T=\SI{40}{ns}$,
confirming its operation within the frequency band of \SI{0.5}{GHz} to \SI{5.5}{GHz} 
(Fig.~\ref{fig:Spara_timegate}(c)).
Additionally, to evaluate the characteristics of the detector IDT,
the transmission characteristics from one chirp IDT to the detector IDT
were also measured (Figs.~\ref{fig:Spara_timegate}(b) and \ref{fig:Spara_timegate}(d)).
Simulation responses shown with gray lines were obtained by calculation based on delta-function
model using ``idtpy''~\cite{idtpy_junliang}.
The simulated results are in good agreement with the experimental results.
The characteristics obtained from this simulation will be used later for reproducing the SAW waveform.

\begin{figure*}
\includegraphics[width=17cm]{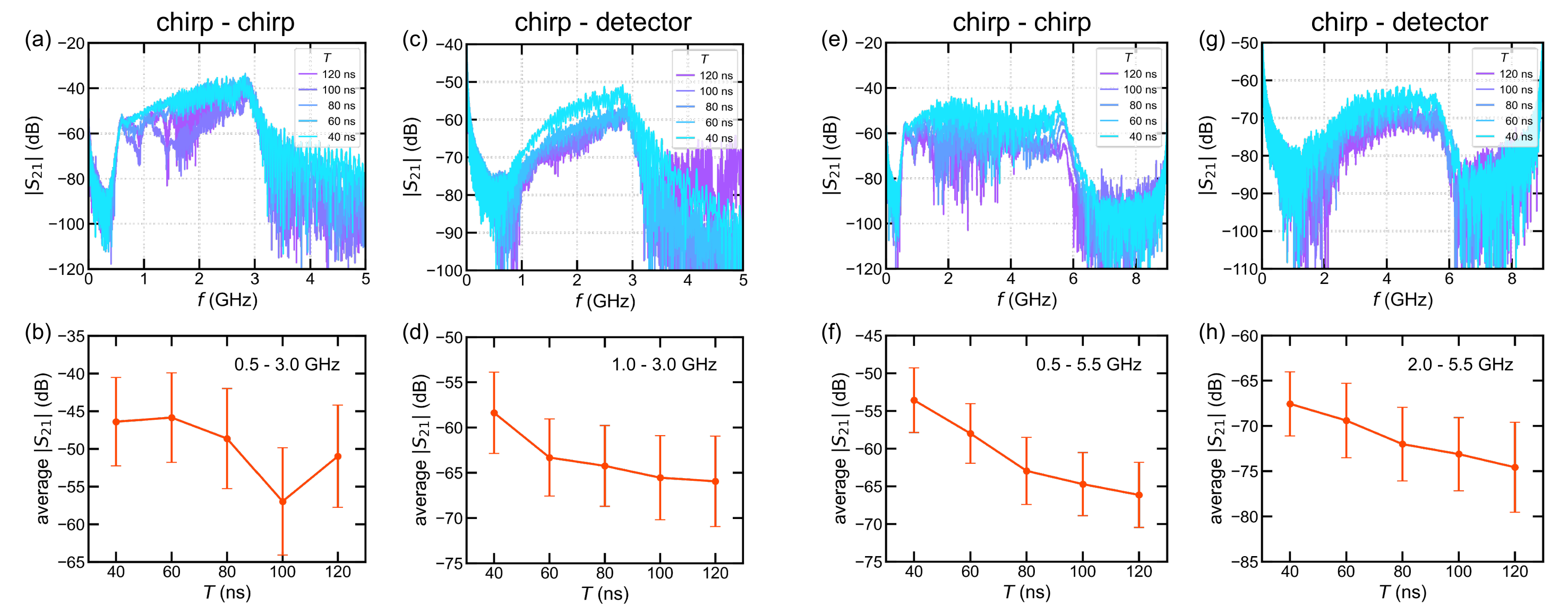}
\caption{\label{fig:Spara_variousT} 
(a) Transmission property $|S_{21}|$ between two chirp IDTs after time gating. Chirp IDTs with different parameters $T = 40,60,80,100,\SI{120}{ns}$ are compared. These devices are designed with common parameters $f_1 = \SI{0.5}{GHz}$, $f_N = \SI{3.0}{GHz}$ and $v_\mathrm{SAW(design)} = \SI{3800}{m/s}$.
(b) Average of $|S_{21}|$ in the range $f= \SI{0.5}{GHz}$ to $f = \SI{3.0}{GHz}$ for various $T$. Error bars indicate standard deviation.
(c) Transmission property $|S_{21}|$ between chirp IDT to detector IDT after time gating. Detector IDTs are regular IDTs with 1.5 pairs of fingers, $f_\mathrm{0d} = \SI{2.5}{GHz}$, $v_\mathrm{SAW(design)} = \SI{3800}{m/s}$.
(d) Average of $|S_{21}|$ in the range $f= \SI{1.0}{GHz}$ to $f = \SI{3.0}{GHz}$ for various $T$. Error bars represent standard deviation.
(e)-(h) The same datasets as (a)-(d) for the devices designed with $f_1 = \SI{0.5}{GHz}$, $f_N = \SI{6.0}{GHz}$. Detector IDTs are regular IDTs with 1.5 pairs of fingers, $f_\mathrm{0d} = \SI{4.5}{GHz}$, $v_\mathrm{SAW(design)} = \SI{3800}{m/s}$.
For the time-gated analysis, a time window of \SI{170}{ns} – \SI{450}{ns} was used for chirp-to-chirp measurements and \SI{80}{ns} – \SI{220}{ns} for chirp-to-detector measurements, applied consistently across all devices at both \SI{3}{GHz} and \SI{5.5}{GHz}.
}
\end{figure*}

In a chirp IDT on GaAs, it is known that an increase in 
$T$ enhances the SAW excitation efficiency.\cite{Ota_thesis}
To investigate the $T$ dependence on a LiNbO$_3$ substrate,
chirp IDTs with different $T$ values were fabricated on the same substrate,
and their characteristics were evaluated.
$|S_{21}|$ between a pair of chirp IDTs for the devices with various $T$ are shown in Fig.~\ref{fig:Spara_variousT}(a).
When $T$ is increased, transmission signals decreased and some dip structures appeared.
We show the average value of $|S_{21}|$ from \SI{0.5}{GHz} to \SI{3.0}{GHz} for various $T$ in Fig.~\ref{fig:Spara_variousT}(b).
The amplitude of transmission tends to decrease with increasing of $T$.
We also measured transmission properties between one chirp IDT and the detectors IDT (Figs.~\ref{fig:Spara_variousT}(c) and \ref{fig:Spara_variousT}(d)).
As the variation in quality of detector IDT is considered to be relatively small,
in this case, the tendency of decrease of $|S_{21}|$ is observed more clearly.
A similar behavior was observed for the \SI{5.5}{GHz} chirp IDT device (Figs.~\ref{fig:Spara_variousT}(e)–\ref{fig:Spara_variousT}(h)).
This is a different tendency from the one for GaAs~\cite{Ota_thesis}.

One possible reason is that the input power is enclosed in the chirp IDT itself
as the electro-mechanical coupling coefficient $K^2$ of LiNbO$_3$ is 80 times higher than GaAs.
The electromechanical coupling coefficient $K^2$ is given by
$K^2 \propto | (v_0 - v_\mathrm{m})/{v_0} |$,
where $v_0$ and $v_\mathrm{m}$ are the SAW velocities when the substrate surface is adjacent to vacuum and
an infinitesimally thin perfect conductor, respectively~\cite{Datta_SAW, 56_Ingebrigtsen_1969}.
A high $K^2$ means a large difference between $v_0$ and $v_\mathrm{m}$.
This leads to larger reflections of the SAW at the boundary between
the free surface and the metal electrode of IDT fingers.
As a result, an increase of the number of fingers will enhance the reflection of SAWs within IDTs more rapidly on LiNbO$_3$ than on GaAs.
Here we comment that this effect is more significant when the width of metal electrode is comparable to or longer than the SAW wavelength.
To reduce the influence of the effect we designed the IDTs such that a SAW with a longer wavelength passes through the metal electrodes to generate a SAW with a shorter wavelength (see Fig.~\ref{fig:dev_pic}(c)).
In a LiNbO$_3$ substrate, a similar tendency has also been observed for an IDT with a single resonant frequency at \SI{4}{GHz}.
The resonant transmission between the IDTs saturates around $T = \SI{50}{ns}$.
For more detailed discussion, further investigation of power confinement effects in chirp IDTs, for example by comparing with narrow-band IDTs of minimal size or by directly measuring surface displacements using optical methods such as laser Doppler vibrometry \cite{Knuuttila_2000, Kokkonen_2008, Hashimoto_2011, Fisicaro_2025}, would provide valuable insights.
In this study, measurements were only conducted with parameters above $T=\SI{40}{ns}$.
To find the optimal condition more investigation within the range of $T < \SI{60}{ns}$ is required.


\begin{figure}
\includegraphics[width=7.8cm]{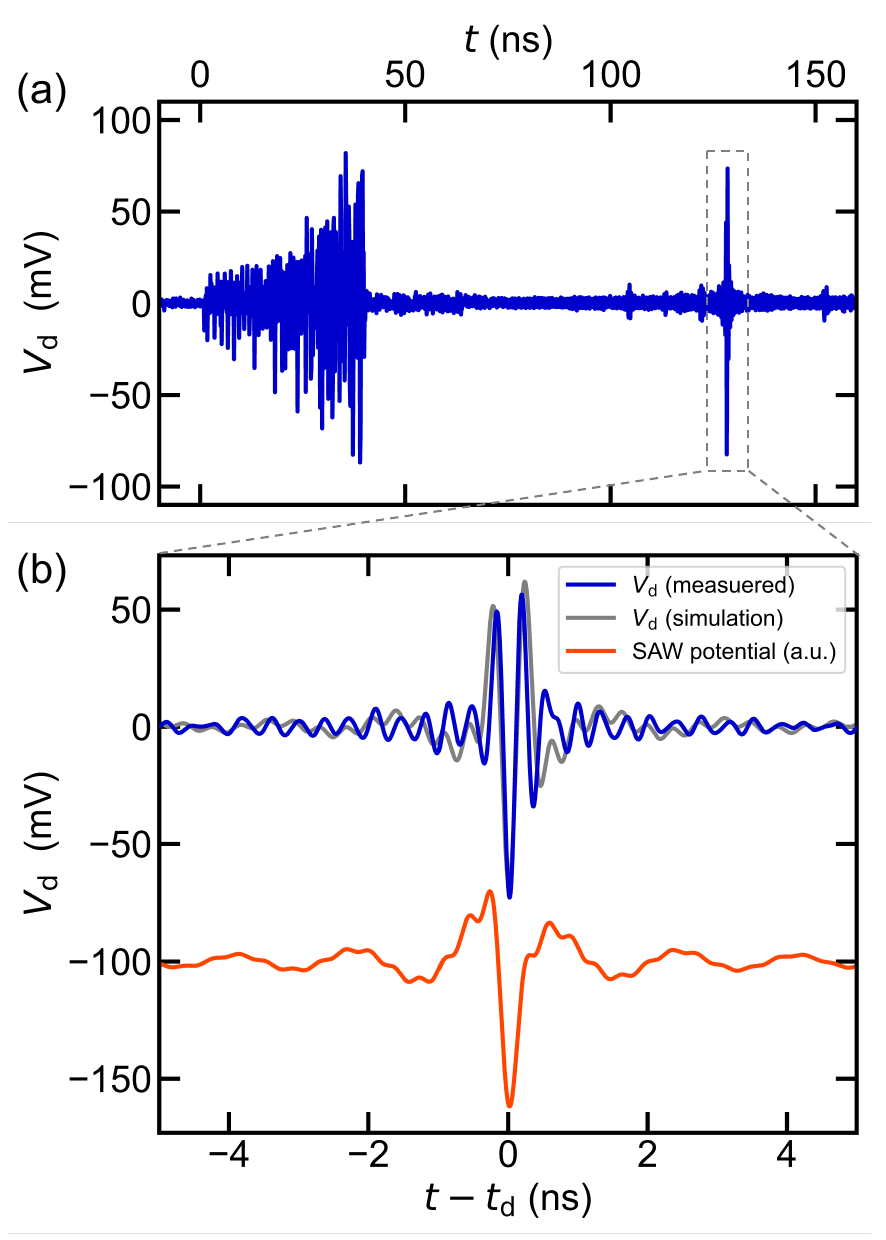}
\caption{\label{fig:realtime} 
Time-resolved measurement of a SAW pulse signals for $T = \SI{40}{ns}$ device. (a) Voltage signals detected by the detector IDT. (b) Zoomed-in area of SAW related signals in $t_\mathrm{d} \approx \SI{128}{ns}$. $t_\mathrm{d}$ is the flight time of the SAW signal between the chirp IDT and the detector IDT. The blue line indicates the measured data and the gray line is the simulated signal detected by the detector IDT. The red line is SAW electric potential obtained by deconvolving the property of the detector IDT from the simulated signal (arbitrary unit).}
\end{figure}

Next we performed time-resolved measurement using the \SI{3}{GHz} chirp IDT device designed with $T = \SI{40}{ns}$, which shows the highest transmission $S_{21}$ from the chirp IDT to the detector IDT (Figs.~\ref{fig:Spara_variousT}(c) and \ref{fig:Spara_variousT}(d)).
From the design parameters of the IDT, $v_\mathrm{SAW(design)} = \SI{3800}{m/s}$, $T = \SI{40}{ns}$, $f_1 = \SI{0.5}{GHz}$, and $f_N = \SI{3}{GHz}$, we obtained the waveform of the input signal using ``idtpy''~\cite{idtpy_junliang}.
The input signal was generated by AWG, amplified by the broadband amplifier (\SI{29}{dB} gain), and applied to the chirp IDT.
The amplitude of the output signal from AWG was set to \SI{0.4}{\volt_{pp}} for all the frequencies.
The SAW signal generated by the chirp IDT propagated along the wafer and was converted to the voltage signal by the detector IDT.
Then it was measured by using a sampling oscilloscope.
In this measurement, it is important to tune the input signal to compensate the mismatch between $v_\mathrm{SAW(design)}$ and the actual SAW velocity, and also the imperfection of the IDT fabrication such as the metallization ratio of the IDT fingers.
The SAW velocity also depends on temperatures.
In this study, all measurements were performed at room temperature.
However, when measurements are carried out at low temperatures, 
the input signal needs to be optimized at each temperature.

The measured signal is shown in Fig.~\ref{fig:realtime}(a).
The signal appearing in a shorter time ($t < \SI{40}{ns}$) represents
an electromagnetic crosstalk propagating from the chirp IDT
with the speed of light during the SAW excitation.
The SAW signal was observed at $t = t_\mathrm{d} \approx \SI{128}{ns}$, 
where $t_\mathrm{d}$ is the flight time of the SAW signal between the chirp IDT and the detector IDT.
The measured SAW signal is shown in Fig.~\ref{fig:realtime}(b) with the blue line.
Here this signal differs from the actual waveform of the SAW
due to the finite bandwidth of the detector IDT.
The actual waveform can be numerically calculated based on impulse-response model using ``idtpy''~\cite{idtpy_junliang}.
To do that we firstly reproduced the measured signal in the simulation, which is shown with the gray line.
After finding the good agreement between the measured signal and the simulation,
we subtract the influence of the detector IDT from the simulation,
obtaining the actual SAW waveform shown with the red line in Fig.~\ref{fig:realtime}.
Consequently, a SAW pulse with a FWHM of approximately \SI{0.3}{ns} was obtained.
Its shape represents the time-dependent electric potential of the SAW pulse generated by the chirp IDT.
Note that the displacement of the piezoelectric substrate is proportional to the electric potential induced by the SAWs.
We also note that our time-resolved measurement detects the average shape of the SAWs over the aperture and is insensitive to the transverse amplitude distribution across the aperture.
The transverse amplitude distribution can affect the material property modulations, depending on parameters such as the IDT’s aperture length, SAW wavelength, and the material’s position relative to the IDT. Special IDT designs, such as apodization \cite{Omori_2011, Yin_2022, Guo_2023}, are useful for suppressing the transverse amplitude distribution.

The detected voltage amplitude $V_\mathrm{d}$ was approximately \SI{130}{m \volt_{pp}} (peak to peak).
When we compare the power conversion efficiency of this chirp IDT with
the similar one ($T=\SI{120}{ns}$, $f_1 = \SI{0.5}{GHz}$, $f_N = \SI{3.0}{GHz}$)
on a GaAs substrate in a previous study~\cite{Ota_thesis},
it is 45 times more efficient for the chirp IDT on LiNbO$_3$.
In this study, we focused only on the generation of a single-cycle SAW pulse.
A chirp IDT, however, can superimpose frequency components from $f_1$ to $f_N$
by exciting a waveform with frequency $f(t)$ at a given time $t$.
By appropriately controlling the amplitude and phase
of each frequency component, it is possible to excite SAWs with arbitrary waveforms
that can be expressed within the frequency band.

In conclusion, we have demonstrated creation of a solitude-SAW pulse on LiNbO$_3$ substrate.
We fabricated chirp IDT devices on a 128$^{\circ}$Y-cut LiNbO$_3$ substrate, ranging from 0.5 to \SI{5.5}{GHz}.
We conducted time-resolved measurement using a chirp IDT with a frequency range of \SI{0.5}{GHz}–\SI{3.0}{GHz} and $T = \SI{40}{ns}$, confirming the generation of a SAW pulse with a FWHM of approximately \SI{0.3}{ns}.
From the frequency response of the devices with the different numbers of IDT finger pairs in the range of $T = \SI{40}{ns}$ to \SI{120}{ns}, we found that the amplitude of transmission signal decreases when we increase the number of IDT finger pairs.
This tendency is in contrast to the similar chirp IDT on GaAs substrate ($f_1 = \SI{0.5}{GHz}$, $f_N = \SI{3}{GHz}$), where the amplitude increases while $T$ increases from \SI{40}{ns} to \SI{120}{ns}.
Nonetheless, with a smaller number of finger pairs the power conversion efficiency of a chirp IDT on LiNbO$_3$ is 45 times larger than the one on GaAs having more finger pairs.

The efficient generation of the SAW pulse demonstrated in this study will open up possibilities in various thin-film material investigations.
In spintronics, for example, generation of pulsed spin current or a single skyrmion, and manipulation of them can be envisioned.
Furthermore, combination of the demonstrated efficient SAW pulse generation with thin-film materials makes it possible to modulate the material properties in time-resolved manner.
Studying the dynamics of material-property modulations or quasiparticle excitation will expand the research associated with thin-film materials.

This work was supported by JSPS KAKENHI (Grant 
Nos. JP23H00257, JP22KJ2180) and 
JST FOREST (Grant No. JPMJFR2134).

\section*{Author Declarations}
\subsection*{Conflict of Interest}
The authors have no conflicts to disclose.

\subsection*{Author Contributions}
\textbf{Koji Fujiwara:} Conceptualization (equal); Data curation (lead); Formal analysis (lead); Investigation (lead); Methodology (equal); Software (equal); Validation (equal); Visualization (lead); Writing – original draft (lead); Writing – review \& editing (equal).
\textbf{Shunsuke Ota:} Formal analysis (supporting); Investigation (supporting); Methodology (equal); Software (equal); Writing – review \& editing (equal).
\textbf{Tetsuo Kodera:} Writing – review \& editing (equal).
\textbf{Yuma Okazaki:} Methodology (equal); Writing – review \& editing (equal).
\textbf{Nobu-Hisa Kaneko:} Resources (equal); Writing – review \& editing (equal).
\textbf{Nan Jiang:} Writing – review \& editing (equal).
\textbf{Yasuhiro Niimi:} Conceptualization (equal); Funding acquisition (equal); Resources (equal); Writing – review \& editing (equal).
\textbf{Shintaro Takada:} Conceptualization (equal); Formal analysis (supporting); Funding acquisition (equal); Investigation (supporting); Methodology (equal); Project administration (lead); Resources (equal); Supervision (equal); Software (equal); Supervision (equal); Validation (equal); Visualization (supporting); Writing – original draft (supporting); Writing – review \& editing (equal).

\section*{Data Availability}
The data that support the findings of this study are available
from the corresponding author upon request.

\section*{REFERENCES}
\nocite{*}
%

\end{document}